\documentclass[11pt,a4]{article}
\usepackage{geometry}                
\usepackage{graphicx}
\usepackage{amssymb}
\usepackage{epstopdf}
\usepackage{hyperref}
\usepackage{setspace}

\pdfoutput=1

\begin{document}

\title{The (A)temporal Emergence of Spacetime\thanks{We thank Baptiste Le Bihan, Tushar Menon, Daniele Oriti, and James Read for comments and discussions, as well as audiences at the PSA in Atlanta, the BSPS in Edinburgh, in Rome and at SISSA. This work was performed under a collaborative agreement between the University of Illinois at Chicago and the University of Geneva and made possible by grant number 56314 from the John Templeton Foundation. Its contents are solely the responsibility of the authors and do not necessarily represent the official views of the John Templeton Foundation.}}
\author{Nick Huggett, University of Illinois at Chicago\\
Christian W\"uthrich, University of Geneva}
\date{31 October 2017}
\maketitle

\begin{center}
Forthcoming in {\em Philosophy of Science}.
\end{center}

\begin{abstract}\noindent
This paper examines two cosmological models of quantum gravity (from string theory and loop quantum gravity) to investigate the foundational and conceptual issues arising from quantum treatments of the big bang. While the classical singularity is erased, the quantum evolution that replaces it may not correspond to classical spacetime: it may instead be a non-spatiotemporal region, which somehow transitions to a spatiotemporal state. The different kinds of transition involved are partially characterized, the concept of a physical transition without time is investigated, and the problem of empirical incoherence for regions without spacetime is discussed.
\end{abstract}

\section{Introduction}\label{sec:intro}

An ordinary liquid can be considered a derived, `effective' entity, as it is composed of molecules. Knowing the relationship between molecular and hydrodynamical descriptions helps us understand the nature of both molecules and liquids. But a collection of molecules is not always a liquid: they might instead be in a state in the solid phase. So one also wants to understand transitions from a state with an effective solid description to one with an effective hydrodynamical description, the process of melting at the molecular level. This paper addresses a parallel question for spacetime in quantum gravity.

Many proposed theories of quantum gravity (QG) suggest that their fundamental quantities and structures do not include all the familiar ones of theories involving classical spacetimes (especially the relativistic spacetimes of general relativity (GR) and quantum field theory). If so, relativistic spacetime must to some extent be a derived, effective entity, and not all fundamental states need correspond to full classical spacetime: some might not have an effective spacetime description. In turn then, spacetime might not only be composed of something non-spatiotemporal, QG may permit a `transition' from a non-spatiotemporal to a spatiotemporal phase, an event maybe to be identified with the big bang, with `earlier' non-spatiotemporal states of the universe.

Let us be careful in several ways from the start. First, as the scare quotes indicate, it does not make obvious literal sense to talk of a `transition' from the non-temporal to the temporal, since transition is a temporal notion. We will discuss this issue later. Second, for brevity, we will say `non-spatiotemporal', but we mean, more carefully, `less than fully spatiotemporal in some significant regard'.  Finally, while `big bang' strictly refers to an initial singularity in classical GR, QG aspires to  `smooth over' any singular behaviour with continuous quantum behaviour. Likely, classical behaviour becomes a poor description at some finite time from the singularity, and so any `big bang' transition is not happening punctually at the singularity, but in some way around it. (The potential difficulty is that the very notion of time itself may break down, making it impossible to ask when or how long a transition occurs, beyond a certain classical early time.)

A scenario in which the universe makes such a phase transition from the non-spatiotemporal to the spatiotemporal has been claimed by Oriti (2014), in his discussions of `geometrogenesis'. Here we discuss two further examples (which arguably do not amount to geometrogenesis) to investigate whether and in what way spacetime might come into being at the big bang.\footnote{See Vaas 2004 for an earlier discussion.} In \S\ref{sec:string}, we introduce a model of cosmology based on string theory. In \S\ref{sec:lqc}, we turn to cosmological models based on Loop Quantum Gravity (LQG), thus covering the two main approaches to QG. Finally, \S\ref{sec:conc} addresses three conceptual issues that arise in such models: how should we categorize the emergence (or non-emergence) of spacetime in these contexts, how can we conceive of a `transition' from timelessness, and how do these cosmological models relate to the problem of `empirical incoherence'?

\section{String Quantum Cosmology}\label{sec:string}

Huggett and Vistarini (2015) discussed the derivation of GR from string theory.  In the simplest case, the classical string action is simply the spacetime area of the string worldsheet: a quantity clearly depending on the Lorentzian metric, $g_{\mu\nu}$, of the background spacetime. This area is invariant under any internal reconfiguration (`stretching') of the string that leaves unchanged the spacetime surface it occupies, so such `Weyl' (local conformal) transformations will be symmetries of the classical action.\footnote{Strings are unlike non-relativistic, Hooke's Law springs in this regard, precisely because the action is a relativistic invariant.} On quantization, local symmetries must be preserved on pain of non-unitarity; but it has been shown for this action (Callan et al.\ 1985) that a necessary condition for quantum Weyl symmetry is that the background metric satisfy the Einstein vacuum field equations, $R=0$, the condition of Ricci (scalar) flatness. And the result generalizes: if other background fields (fermionic matter, or bosonic gauge fields, for instance) appear in the string action, then Weyl symmetry requires that they mutually satisfy the appropriate Einstein field equations. 

For instance, if a scalar field, $\phi(x)$, known as the `dilaton', is added to the string action, then parallel reasoning leads to the following pair of equations (Gasperini 2007, \S2.1):
\begin{equation}
\label{eq:SEFE}
G_{\mu\nu} +\nabla_\mu\nabla_\nu\phi + 1/2g_{\mu\nu}(\nabla\phi)^2 - g_{\mu\nu}\nabla^2\phi = \lambda_s^{D-2}e^\phi T_{\mu\nu}
\end{equation}
and
\begin{equation}
\label{eq:dil}
R + 2\nabla^2\phi -(\nabla\phi)^2 = \lambda_s^{D-2}e^\phi\sigma.
\end{equation}
$G_{\mu\nu}$ is the usual Einstein tensor, $\lambda_s$ is the characteristic string length (the fundamental free parameter of the theory), $T_{\mu\nu}$ is the stress-energy tensor of all other matter fields, and $\sigma$ is the matter `charge' to which the dilaton couples. (\ref{eq:SEFE}) is a modified form of the standard field equations, in which (a) the dilaton appears with $g_{\mu\nu}$, and (b) the coupling strength of the gravitational field is not the gravitational constant, but depends on the dilaton as $\lambda_s^{D-2}e^\phi$. Huggett and Vistarini (2015) discussed how the dilaton can control the dimensions of spacetime. Here we also see that interaction strengths depend on $\phi$: in this case the strength of gravity---but also the string coupling, $g_s$, which controls the rate at which strings split and join.

Two points should be emphasized. First, these so-called `background fields', including the spacetime metric and dilaton, are not (despite their name) distinct classical fields, but instead represent the effects of special string states. In particular, gravitons and dilatons are in the string particle spectrum, and the corresponding terms in the action are meant to represent the effects of coherent states of such stringy quanta (Green et al 1987; \S3.4).\footnote{Formally, at least,  $g_{\mu\nu}$ represents the effect of a graviton field in Minkowski spacetime.}

Second, string theory involves a double perturbation expansion, both with respect to $g_{\mu\nu}$,\footnote{$g_{\mu\nu}$ formally acts like a field coupling in the action, whose strength depends on the curvature. Formally, perturbative expansions are carried out in powers of a small fundamental constant $\alpha'=\sqrt{\lambda_s}$.} and with respect to $g_s$. The former represents the effects of the extended, stringy nature of the graviton; the latter represents the quantum effects of strings splitting and joining. Neither parameter is a constant, but depend on the background curvature and dilaton respectively: only if both are weak is string theory in a perturbative sector. Indeed, the effective equations (\ref{eq:SEFE}-\ref{eq:dil}) are derived at the lowest, tree level, and so break down when higher order corrections are significant, or perturbation theory fails altogether.

The equations can be simplified by assuming a homogeneous, isotropic, flat, matter-free spacetime\footnote{The following example and its implications are drawn from Gasperini (2007), especially \S4.1.}. In this case $T_{\mu\nu}=\sigma=0$, while
\begin{equation}
\label{eq:ametric}
g_{\mu\nu} = \mathrm{diag}\big(1, -a(t)^2, -a(t)^2, -a(t)^2\big),
\end{equation}
so that the metric has a single time-dependent parameter, the `scale factor' $a(t)$, which describes the `relative size' of space, in a  way familiar from GR cosmology. Additionally, we adopt string units, which set $\lambda_s=1/\sqrt{2}$, and introduce the `Hubble parameter', $H(t)\equiv\dot{a}(t)/a(t)$. Then the equations take the form 
\begin{eqnarray}
\dot{\phi}^2 - 6\dot{\phi}H + 6H^2   =   0,\qquad\qquad\dot{H}-H\dot{\phi}+3H^2   =   0, \\
\mathrm{and}  \qquad 2(\ddot{\phi} + 3\dot{\phi}H) - \dot{\phi}^2 - 6\dot{H}-12H^2   =   0.
\end{eqnarray}
They represent the temporal and spatial parts of (\ref{eq:SEFE}), and (\ref{eq:dil}), respectively.

As well as time reversal symmetry ($t\to-t$), these equations also exhibit an interesting `scale inversion symmetry': they are unchanged under

\begin{equation}
\label{ }
a(t)\to 1/a(t) \quad \mathrm{and} \quad \phi(t) \to \phi(t) - 6\ln a(t).
\end{equation}
This symmetry is not an artifact of the special case, but hold of (\ref{eq:SEFE}-\ref{eq:dil}); indeed, it is a special case of a more general symmetry for strings with background fields (Gasperini 2007, \S4.3).

We flag this fact for two reasons. First, as we will illustrate below, this pair of symmetries provides a general method for constructing pre-big bang epochs from post big bang solutions. Second, scale inversion is analogous to T-duality (Huggett 2017) in spaces that are neither time-independent, nor compact (in the case of our effective classical theory). We therefore commend it for philosophical attention.

Our equations of motion have a solution for $t>0$
\begin{equation}
\label{eq:t>0}
a(t) = (t/t_0)^{1/\sqrt{3}} \qquad H(t) = 1/(\sqrt{3}t) \qquad \phi(t) = (\sqrt{3}-1)\ln(t/t_0).
\end{equation}
These describe an expanding ($\dot{a}>0$), but decelerating ($\ddot{a}<0$) Friedman-Lema\^{\i}tre-Robertson-Walker spacetime (with $\Lambda=0$). Moreover, since the Hubble parameter provides an indication of the spacetime curvature, we see that as $t\to0$ there is a curvature singularity---an initial `big bang'.  

Applying time and scale inversion symmetries there is another solution, for $t<0$
\begin{equation}
\label{eq:t<0}
a(t) = (-t/t_0)^{-1/\sqrt{3}} \qquad H(t) = -1/(\sqrt{3}t) \qquad \phi(t) = -(\sqrt{3}+1)\ln(-t/t_0),
\end{equation}
a spacetime which is expanding ($\dot{a}>0$), and accelerating ($\ddot{a}>0$)---because of the dilaton, this solution exhibits inflation. It too has a curvature singularity at $t\to0$, however, as $t\to-\infty$ the curvature vanishes: this solution represents a spacetime which has a perfectly continuous past (not an early big bang), but which has a \emph{final} singularity. 

Turning to the dilaton field, we see that $\phi$ diverges as $t\to0$ in both solutions---with opposite sign. It is this divergence that drives the early universe to the big bang singularity in the model. Now, recalling that the string coupling $g_s\sim\exp\phi$, we see that the first solution has $g_s\to0$ at $t\to+0$, while the second solution has $g_s\to\infty$ at $t\to-0$: for the latter but not the former, the string perturbative expansion fails at $t\to0$. However, since the curvature is singular at $t\to0$ in both solutions, the $g_{\mu\nu}$ expansion fails at $t\to0$ for both. Since our equations are perturbative, they thus fail at the big bang.
  
Given these solutions for epochs of positive and negative times, which agree on the geometry as $t\to0$, one naturally wonders whether they can be meaningfully joined together at $t=0$, a big bang that occurs between an infinite past and future? Since the geometries are singular at $t=0$, this proposal does not strictly make sense, but recall that the equations of motion which produce these solutions are themselves perturbative, and only apply for weak curvature and string coupling. Hence, as we just saw, we cannot trust them too close to the curvature singularity, and the question becomes whether quantum and stringy corrections will prevent the singularity, and allow a consistent matching of the solutions: that is, is there a complete string theoretical model, for which our solutions are early and late time approximations? If so, \emph{what kind of physics do they entail for the big bang}? For if the classical approximation fails, there opens up the possibility that what is left is quantum spacetime!

Here things are, unfortunately, unclear (Gasperini 2007, Chapter 6). Various possibilities arise, and it is not known which holds. Assuming that indeed there is a stringy model connecting the solutions, the following are the most relevant for the purposes of this paper. First, the equations studied are only to lowest order in perturbation theory. As the curvature grows as $t\to0$, higher order corrections in $g_{\mu\nu}$ become significant; perhaps including them causes (\ref{eq:t>0}-\ref{eq:t<0}) to lose their singular behaviour, so the classical spacetime approximation remains valid throughout. Put another way, the correct effective theory around $t=0$ is a modified form of GR, in which a big bang singularity is avoided. According to Gasperini (2007, 230) there are indications that this scenario does not hold.

Second, it is possible that the growth of the dilaton, and hence $g_s$, for $t<0$ make higher order string corrections important, though the perturbative scheme still holds through the transition to the classical $t>0$ solution (at which point the first order solution again suffices). Since such corrections represent (inter alia) the creation and annihilation of stringy gravitons, in this scenario the effective classical description of gravity breaks down, and is replaced by a quantum theory of superpositions of classical gravitational fields. Some modeling of this scenario exists, but no detailed smooth transition between the two classical solutions is known (Gasperini \S6.3, Appendix 6A).\footnote{Some models explain why only three of the nine spatial dimensions of supersymmetric string theory are large.}

Finally, it is possible that the growth of curvature and dilaton as $t\to0$ cause perturbative string theory to fail altogether, and an exact, `M-theory' is needed to describe the physics. Of course, this theory is elusive: neither its fundamental quantities nor equations are known. However, especially because of the many dualities linking its (assumed) limits, it is widely believed by string theorists that the fundamental degrees of freedom are non-spatiotemporal. Supposing that (and supposing that our classical solutions are stringy phases of a single M-theoretic model), then the region around $t=0$ at which the solutions join, is fundamentally non-spatiotemporal: not well-described by a classical spacetime, nor even by a quantum one.

\section{Loop Quantum Cosmology}\label{sec:lqc}

Loop Quantum Gravity (LQG) proceeds by a canonical quantization of GR. As such, it casts GR as a Hamiltonian system with constraints. All but one of the constraint equations have been solved: the Hamiltonian constraint equation, in loose analogy to the standard case the `dynamical' equation of the theory, has not so far succumbed to solution. Physicists have pursued two work-arounds. Some have attempted to replace the canonical `dynamics' with a supposedly equivalent covariant one (Rovelli and Vidotto, 2015). Others have instead studied simplified systems with the original canonical `dynamics'. 

Loop Quantum Cosmology (LQC) undertakes the latter: already at the classical level, it imposes a symmetry reduction on the space of admissible models.\footnote{For an accessible, but technical introduction to LQC, see Bojowald (2011).} We start with LQG, where the geometry is not expressed in terms of metric variables, but instead the canonical variables are a connection and a (densitized) triad. These basic variables are then used to construct a `holonomy-flux algebra'. In  full LQG, one arrives at the `kinematical Hilbert space' $\mathcal{H}_K$---the space of those states satisfying all but the Hamiltonian constraint. The so-called `spin network states' form an orthonormal basis in $\mathcal{H}_K$; these states are eigenstates of the `area' and `volume' operators. The spin network states are constructed from a `quantum geometrical vacuum' state using the holonomies as `creation' operators, raising the excitation level of the fluxes. They can be represented as graphs labeled with spin representations on their edges and vertices. Since the `dynamics' is not yet accounted for, the spin network states are routinely interpreted as `spatial' such that physical space is a quantum superposition of spin network states with well-behaved geometric properties.

As is usual in cosmology, LQC assumes spatial isotropy, such that the spatial geometry is given by just one degree of freedom: the  scale factor $a$ of (\ref{eq:ametric}). It is in this sense that LQC addresses the `cosmological' sector of LQG. Thus, all degrees of freedom are `frozen', except for one (or three, in the anisotropic case). This symmetry reduction is already imposed at the classical level, prior to quantization.\footnote{It is unknown whether quantizing first and symmetry-reducing second will in fact deliver an equivalent result, although in principle it should.} Unlike in the full theory, where in general the spin networks are complicated and irregular, the isotropic quantum configuration is highly regular and can be represented by a lattice graph consisting of straight edges of roughly the same basic length and similar-sized surfaces of an area roughly the square of the basic length. 

More concretely, the single remaining degree of freedom is captured by the `scale factor' operator $\hat{p}$, essentially corresponding to the classical scale factor $a$ via $a = \sqrt{|p|}$, where the $p$'s are the eigenvalues of $\hat{p}$. The scale factor operator has a (quasi-)discrete spectrum, which includes zero\footnote{The spectrum is $\mathbb{R}$, but the eigenstates are normalizable, and thus the spectrum is discrete by definition.}: the isotropic spatial quantum geometry is `atomic' in this sense. In order to obtain the (symmetry-restricted) physical Hilbert space $\mathcal{H}^s$, one expresses and then solves the simplified Hamiltonian constraint equation on the basis of a momentum representation of states $|\Psi\rangle$ in $\mathcal{H}^s$, using $|\Psi\rangle = \Sigma_{\mu} \psi (\phi, \mu) |\mu\rangle$, where $\phi$ designates matter fields and the $|\mu\rangle$ form a complete basis of eigenstates of $\hat{p}$.  Still at the kinematical level, one can build an inverse triad operator, call it $\hat{M}$, capturing the scalar curvature of the isotropic quantum geometry. Classically, the corresponding curvature $a^{-1}$ would diverge for $a\to 0$, as the big bang is approached. But in LQC, $\hat{M}$ turns out to not only have the appropriate classical limit for sufficiently large $\mu$, and hence $a$, but also to exhibit bounded small-$\mu$ behavior, with a peak around a small, but non-zero $\mu$. In fact, at $\mu = 0$, the eigenvalue of the operator is zero. Thus, the curvature singularity at what is classically the big bang vanishes in LQC. 

So far, however, only the `kinematical', or spatial, states are considered. In order to turn LQC into a theory of QG, the physical states must be identified, i.e.\ those states which satisfy the `dynamics', and hence are fully spatiotemporal. Mathematically, the dynamics is implemented in the Hamiltonian constraint equation $\hat{C} |\Psi\rangle = 0$, where whatever $|\Psi\rangle$ satisfy this constraint (and the others) are the physical states we are looking for. These physical states can be expanded in terms of eigenstates of some operator, such as of a scalar field or of the volume, which does not need to be an observable (and so the eigenstates may not themselves be `physical'). Since $|\Psi\rangle$ is extended in `time', the family of these eigenstates can then be interpreted as an `evolving quantum geometry', with the evolution occurring with respect to the degree of freedom captured by the operator---the `internal time': the `variation' of $\Psi$ with respect to this basis stands in for the temporal variation of basis states over the full interval of eigenvalues. All of this can in principle be done in full LQG. It is at this point that the main advantage of LQC over the full theory manifests itself: unlike in the full theory, it is possible to concretely construct the Hamiltonian constraint operator, nurturing the hope that we can grasp important qualitative features of full LQG. 

As it turns out, if we express the general state in the triad eigenbasis $|\mu\rangle$ of $\hat{p}$, and so use the scale factor which essentially corresponds to $\hat{p}$ as an internal time, the Hamiltonian constraint equation becomes a \emph{difference} equation, rather than a differential equation. Essentially, it has the following form:
\begin{equation}
\label{eq:hamcon}
V_+ \psi (\phi, \mu + 1) + V_0 \psi (\phi, \mu) + V_- \psi(\phi, \mu -1) = \hat{H}_{m} \psi (\phi, \mu), 
\end{equation}
where the $V$'s are coefficients making sure that the appropriate classical limit is obtained and $\hat{H}_{m}$ is the matter Hamiltonian. In other words, for a state $|\Psi\rangle$ to qualify as physical, its components in terms of the triad eigenbasis have to satisfy (\ref{eq:hamcon}). In order to determine $|\Psi\rangle$ through (\ref{eq:hamcon}) by means of the $\psi(\phi, \mu)$, the values of $\psi(\phi, \mu)$ must be given for a closed unit interval containing some $\mu\in \mathbb{R}$---the `initial values'. As long as the coefficients do not vanish, the complete state can then be determined recursively through (\ref{eq:hamcon}), including those parts to the other side of the `big bang' at $\mu=0$ from the initially given unit interval. As this is in general possible, the dynamical singularity present in classical cosmology, where nothing can `evolve' through the big bang, is resolved.\footnote{See, however, W\"uthrich (2006) for some caveats concerning the claimed singularity resolution.} 

Although problematic, one reading of the simplified Hamiltonian constraint equation is thus as an evolution equation, with the scale factor as `time variable'. Unlike the case of string cosmology, in which the scale factor was a \emph{function} of the time parameter, it is interpreted here as \emph{(cosmic) time itself}, although some other degree of freedom such as a scalar field could play the role of internal time. Under this interpretation of $\mu$ as `time' since $\mu$ runs the whole gamut of $\mathbb{R}$, a `backward' evolution brings us into a `mirror world' of negative $\mu$. How should we think of this mirror world, and how does it relate to the physics of the more homely late cosmic times? Can we meaningfully speak of what happened `before the big bang'?

There are really two separate and distinct questions. First, what is the physics of the realm `beyond the big bang'? Second, how does that physics connect to the physics of our realm? As for the first point, the `trans'-big-bang physics is a precise mirror image of the `cis'-big-bang physics, except that the spatial orientation is inverted (as indicated by the negative sign). The usual interpretation, just as for the string-theoretic model  in \S\ref{sec:string}, is that we are here faced with the physics of a collapsing universe, which heats up as it contracts, reaches a point of maximum heat and minimum size (zero, in fact), and then rapidly re-expands and cools down. We will offer an alternative interpretation in the next section.

Regarding the second question, since what is quantized in LQC is  a geometric degree of freedom (the scale factor), one might anticipate that the physics of the region joining the two parts of the universe are at most non-spatiotemporal in a very mild, quantum-fuzzy way. However, these intermediate states are not semi-classical, and hence have no intuitively spatiotemporal interpretation. In fact, it appears (Barrau and Grain, 2016; Brahma 2017) as if the same mechanism that is responsible for the resolution of singularities in LQC also leads to a so-called `signature change' in these same models: going backwards in time, from a structure which well approximates a spacetime of Lorentzian signature at late `times' (i.e.\ large values of $\mu$), to a structure with Euclidean signature in the deep quantum regime of early `times' (small values of $\mu$), and then back to Lorentzian trans-big-bang physics. The structure in the deep quantum regime is thus, though geometrical, purely spatial. Thus, there is no connected physical time through the big bang epoch---in fact, there appears to be no time at all---and the interpretation of the scale factor as physical time remains limited. It seems as if our current temporal era emerges, temporally, from an atemporal epoch of the universe.

\section{Discussion and Conclusions}\label{sec:conc}

Moving to quantum theories more fundamental than GR seems to lead us, at least potentially, to \emph{atemporal} (or \emph{aspatiotemporal}) structures. What these fundamental less-than-fully spatiotemporal degrees of freedom are and how they can form spacetime depends on the approach to QG taken. But as we have noted, they raise the possibility that there is a \emph{transition}, as it were, from an `earlier' quantum state, which in general lacks any correspondence to a classical emergent state, to a `later' classical cosmology well described by relativistic spacetime. Indeed, we have seen that just such a phenomenon is at least possible in cosmological models based on string theory and LQG. We want to raise three foundational issues of particular note in the context of these two examples, with the second dividing into three sub-points.

(1) Oriti (2015) offers a hierarchy of levels of emergence, corresponding to the severity of the reconstructions required to recover classical, continuum spacetime from the fundamental structure: the higher the level, the more drastic they are, and correspondingly, the less spatiotemporal the fundamental structure. At the first, lowest, level, we just have the quantum version of otherwise ordinary spatiotemporal, geometrical, degrees of freedom: a graviton field for example. At the next highest level, we find fundamental pre-geometric degrees of freedom, which may still have some spatiotemporal properties, but which may or may not have an effective spacetime description, depending on the state. At yet a higher level, we find fundamental degrees of freedom which do not exhibit any spatiotemporal properties, and which will only come to form something like spacetime, approximately, as a result of some `dynamical process'. As an example at this level, Oriti offers his own approach to quantum gravity called `Group Field Theory', a second quantization of LQG, in which spacetime emerges from fundamental degrees of freedom by a `process' called `geometrogenesis' (cf.\ Oriti 2014). Here, the relevant dynamical process is obviously not a temporal happening; instead, it can be interpreted as a transition point in a phase diagram.
 
These levels are a useful general scheme for organizing kinds of emergence, but our examples reveal some complications. First, our string example, shows how the fate of spacetime hangs on what happens to the perturbative expansion. In general, it is expected that a full theory---M-theory, that is---will involve non-spatiotemporal degrees of freedom, and hence belong at one of the higher levels of Oriti's scheme. However, in the specific model discussed, that is only one possibility: it is also possible that only quantum corrections to gravity are relevant, in which case the model corresponds to one of the lower levels. So the example indicates that it makes sense to distinguish models of a single theory according to their level of emergence, as well as different theories.

In LQC, spacetime gets directly quantized and thus replaced by a different structure. In our example the fundamental structures are regular spin networks, which approximate the classical cosmological models very well at late stages. Their states are discrete structures, whose degrees of freedom, arguably, are still geometric; so in this sense, the disappearance of spacetime remains comparatively mild in this case, even though these are quantum structures, and general physical states are superpositions of spin network states. As geometric as the degrees of freedom may remain, however, the fact that the spacetime signature changes from Lorentzian to Euclidean for sufficiently early times suggests that although the structure may afford a geometric, indeed spatial, interpretation, time does not extend into the very early universe. This example shows that there are interestingly different possibilities at the lowest, quantum spacetime level, that arguably should be distinguished in a more fine grained way: quantum time should be distinguished from pure spatiality!
 
(2a) As noted in \S\ref{sec:lqc}, there are at least two ways to conceive of the LQC model. The standard interpretation considers it as a universe which first contracts and heats up, goes through an extremely dense and hot phase, and then expands and cools down.\footnote{Note the difference from string cosmology, in which both spacetime regions have a finite scale factor $a$ at all times, and $a$ increases in the same temporal sense at all times---including the singularity at $t=0$. (The big bang curvature singularity shows up in the Hubble parameter, not a vanishing scale factor.)} In this case, physical time would have to run continuously through the deep quantum regime in order to result in one unified physical process, of an earlier state evolving into a later one. But as we saw, a signature change around the big bang means that there is no such connected notion of time. (Nor can the scale factor provide one, since it is non-monotonic at $t=0$.)

Alternatively, one could interpret the model as describing the \emph{twin} birth of two universes from the same primordial quantum womb, which then both expand and cool. In this case there is not one process, but two (similar) processes in which an effective spacetime region emerges; and no connected, persistent physical time running from before to after the big bang is required. The puzzle now is how a single atemporal quantum realm can stand in a `before' relation to \emph{both} quasi-classical, temporal regions.

Consider the general case in which an effective spacetime emerges in a theory of QG. The precise way in which time emerges hangs on details of the fundamental physics and the particular ways in which its degrees of freedom combine. However, one expects the emergence to involve the aggregation of a suitably complex collection of fundamental quantities. The way in which not just time, but its \emph{direction}, emerges is then essentially thermodynamic, or statistical, and hence given, for instance, by an increasing entropy. In this case, however, the direction of time can vary by region (as, for instance, a Boltzmann fluctuation approaches equilibrium symmetrically in time). Therefore, in the LQC model we may find a situation in which `during' the big bang time does not emerge from the fundamental degrees of freedom, while in the quasi-classical regions, the fundamental degrees of freedom (assuming sufficient symmetry) combine such that the local effective time is in the direction \emph{away} from the big bang in both the trans- and the cis-big bang realms.\footnote{We do not say that such a situation \emph{must} obtain. For instance, the entropy gradient need not align with the scale factor gradient, $\dot a$, and entropy can in general decrease in the direction of expansion (see Penrose 1989, Ch.\ 7). However, a desideratum for a theory of QG is that it explain the early low entropy of our universe. Thus if the loop program succeeds, it will determine in what direction the effective arrows of time point. So whether the scenario we postulate obtains depends on details of the theory; details that are, as yet, unknown. Similar remarks could be made for string cosmology.}

(2b) One might still ask whether such an effective direction suffices: there is still the problem of how an atemporal region could ever come \emph{before} a temporal one, since in such a case there is no time that could count as being earlier. The answer is to extrapolate local, directed time beyond its proper domain of applicability: the atemporal region is before the effective temporal region, iff timelike curves in the effective spacetime can be extended to the atemporal region in the past, but not in the future, effective direction. (Or the converse if the atemporal region comes after the temporal region.) In the models we have discussed, the big bang region---even if timeless---may be a past limit relative to local determinations of time's arrow. In this sense it can come `before'. (We  discuss this definition further below.)

(2c) There remains, however, the crucial question of how we can even speak of a `transition from' an atemporal phase: in geometrogenesis, or from an M-theory phase, or from a LQC phase of purely spatial geometry. ``How can there be a change from a state without time?", Zeno might ask. The radical possibility is that a theory of atemporal degrees of freedom requires a completely different way of doing physics, which abolishes such notions. But first, how far will a more conservative strategy take us?  That is, note that there seems no in-principle difficulty with a fundamental theory allowing an effectively temporal region to have an atemporal state as its future limit point: such a model would describe a `dissolution' of time. But in a fundamental, atemporal theory there is no fundamental direction to time, and `dissolution' and `emergence' are identical. That is, since the direction of time is only effective, when speaking in terms of fundamental physics we should think of the big bang as just a limit to time, taken as early or late only by convention. Insofar as the dissolution of time is unproblematic, so is its emergence.

However, this conservative response is incomplete. We did not fully specify the sense in which the big bang region is a `limit' to time, or what it means to extend an effective timelike curve `to' an atemporal region. It will of course depend on theoretical details of the region, but by construction it will not involve the region having any temporal extent: for instance, the region might be Euclidean, or lack any manifold structure. From the point of view of an effective temporal description we have an open timelike curve plus one extra `point', which is in fact a structured object, more complex than the 3-dimensional hypersurfaces found at other times. (For simplicity, ignore any indeterminacy in whether points are temporal or not.) This picture could comprehend a transition from the atemporal to the temporal, if the initial point were also considered a time, from which later states evolve. That would require a dynamics in which the atemporal region was an initial state.

A quite different idea is that some new fundamental parameter is needed to play the role of time in the fundamental dynamics: a parameter that varies across the atemporal region, and agrees with effective time in the quasi-classical regions. The transition would then be with respect to this quantity, not effective time. Or perhaps the radical possibility is correct: the very idea of dynamics---and with it `transition'---must be replaced in quantum gravity. All we will say is that there is a very deep conceptual issue here, in which philosophy has profound bearing on physics.

(3) Huggett and W\"uthrich (2013) raised the question of `empirical incoherence' for quantum gravity: if a theory postulates that fundamentally there are no (or very weak) spatiotemporal structures, doesn't it undermine the very possibility of its having empirical support---for all observations are ultimately local in space and time? There we argued that the solution is to seek derived (or `emergent') structures that behave functionally as spatiotemporal ones: playing the right role in the laws of theories with classical spacetime, for instance.

However, the current examples---in the strongest cases of spacetime dissolution---reveal a gap in our answer. If there is a non-spatiotemporal region according to theory $T$, then in that region there are no derived spatiotemporal structures, and hence no possible observations of it. So is a theory including such a region empirically incoherent, at least regarding that portion of the universe? Again we say `no'; for of course observations are possible in the spatiotemporal region, and so $T$ is capable of empirical support by its own lights, and insofar as it is supported, so are its consequences, including the existence of the non-spatiotemporal region.

That said, the situation raises some interesting features. First, the non-spatiotemporal phase is unobservable in a particularly profound sense; more profoundly than a very small object, say. Second, as a result, the inference of the theory beyond the observable is particularly tenuous, and hence arguably any inference from empirical evidence is unusually weak. That said, the evidential situation is better in the future of an atemporal phase than the past, for it should leave traces (i.e., one can benefit from the usual temporal knowledge asymmetry); indeed, the string model makes predictions for traces in the cosmic background radiation. Third, it is also possible that $T$ allows worlds in which there is never a phase transition, and the universe is entirely non-spatiotemporal; in such a world, the theory would indeed be empirically incoherent---in other words, empirical incoherence is a relation between theories and their models.

\ 

In short, the above examples and discussion demonstrate that there are significant implications for our cosmic conception of time in QG accounts of the origin of spacetime. We have barely scratched the surface.

\end{document}